\documentclass[ nobibnotes, aps, prd,onecolumn,nofootinbib,showpacs]{revtex4}%
\usepackage{doublespace}
\usepackage{amsfonts}
\usepackage{amssymb}
\usepackage{graphicx}
\usepackage{amsmath}%
\begin{document}
\title{QCD radiative correction to pair-annihilation of spin-1 bosonic Dark Matter}
\author{Jae Ho Heo}
\email{jheo1@uic.edu}
\affiliation{Physics Department, University of Illinois at Chicago, Chicago, Illinois
60607, USA }

\begin{abstract}
The next-to-leading order (NLO) QCD corrections are calculated for the
pair-annihilation of spin-1 dark matter (DM) by dimensionally regularizing
both ultraviolet and infrared singularities in non-relativistic limit ($v\ll
1$). The complete $O(\alpha_{s})$ correction is about $8\%$ due to the
massless gluon contribution. An extra $5\%$ will be added if there is a new
interaction from a massive gluon of approximately same mass as the DM
particle. The NLO QCD correction could give the sizable shift to the DM mass
constrained by relic density measurements.

\end{abstract}

\maketitle

\section{Introduction}

The spin-1 dark matter candidate $B^{\prime}$, which is the $Z_{2}$-odd
partner of the hypercharge gauge boson $B$, appears in interesting models like
universal extra dimensions (UED) \cite{tappel} and Little Higgs (LH)
\cite{narkani}, motivated by solving the gauge hierarchy problem. The
phenomenology of spin-1 DM ($B^{\prime}$) has been well studied at the tree
level. With advances in precision measurements in cosmology and astrophysics
as well as the advent of LHC era, the NLO calculation of the $B^{\prime
}B^{\prime}$ annihilation becomes timely for accurate analysis. We present the
next-to-leading order (NLO) QCD corrections for the pair-annihilation of
spin-1 bosonic dark matter by dimensionally regularizing both ultraviolet (UV)
and infrared (IR) singularities in the non-relativistic limit ($v\ll1$). Our
analysis of the process ($B^{\prime}B^{\prime}\rightarrow q\overline{q}$) is
applicable to general spin-1 DM, as in UED, LH, etc. We assume that the quark
field $q$ interacts with its $Z_{2}$-odd partner $\widetilde{q}$ in the form.%

\begin{equation}
\mathcal{L}\supset-g_{Y}\overline{\widetilde{q}}\gamma^{\mu}(\widetilde{Y}%
_{L}P_{L}+\widetilde{Y}_{R}P_{R})qB_{\mu}^{\prime}+h.c.
\end{equation}

The coupling $g_{Y}$ and the number $\widetilde{Y}_{L}$ in the UED would be
the usual hypercharge coupling and the corresponding quantum number,%

\[
g_{Y}=\frac{e}{\cos\theta_{W}},\text{ \ }\widetilde{Y}_{L}(u,d)=\frac{1}%
{6},\text{ \ }\widetilde{Y}_{R}(u)=\frac{2}{3},\text{ \ }\widetilde{Y}%
_{R}(d)=-\frac{1}{3}%
\]

However in the Littlest Higgs model (L$^{2}$H) \cite{narkani}, $\widetilde
{Y}_{L}(u,d)=\frac{1}{10}$ and otherwise $\widetilde{Y}_{R}=0$.

We also assume that the mass $\widetilde{M}$ of $\widetilde{q}$ is not too far
above the mass $M$ of the spin-1 DM particle $B^{\prime}$. The approximate
relation $\widetilde{M}\approx M$ is valid in UED, but a choice of parameters
in other models is required. However, such a choice allows us to carry through
analytical calculation and give simple results.

Fig.1 shows the Born diagrams for $B^{\prime}(p_{1})B^{\prime}(p_{1}^{\prime
})\rightarrow q(p_{2})\overline{q}(p_{2}^{\prime})$. The exchange of the
$Z_{2}$-odd quark line is bold-faced. The amplitude for the Born diagrams is
given by%

\begin{equation}
\mathcal{M}_{B}=-g_{Y}^{2}\widetilde{Y}_{L}^{2}\overline{u}(p_{2})\left(
\gamma^{\mu}\frac{{\not p  }_{2}-{\not p  }_{1}+\widetilde{M}}{t-\widetilde
{M}^{2}}\gamma^{\nu}+\gamma^{\nu}\frac{{\not p  }_{2}-{\not p  }_{1}^{\prime
}+\widetilde{M}}{u-\widetilde{M}^{2}}\gamma^{\mu}\right)  P_{L}v(p_{2}%
^{\prime})\epsilon_{\mu}(p_{1})\epsilon_{\nu}(p_{1}^{\prime})\text{\ }%
+\text{RH}%
\end{equation}
%

\begin{figure}
[ptb]
\begin{center}
\includegraphics[
trim=0.000000in 0.000000in -0.080171in 0.000000in,
height=2.179cm,
width=5.0215cm
]%
{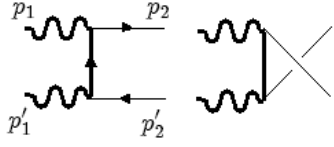}%
\caption{The Born diagrams for $B^{\prime}B^{\prime}\rightarrow q\overline{q}%
$. The bold wavy lines represent the DM candidate gauge bosons ($B^{\prime}$),
the bold solid lines represent the heavy quarks (the $Z_{2}$-odd partners of
the outgoing quarks) and the light solid lines correspond to the quarks.}%
\label{fig1}%
\end{center}
\end{figure}

In the extremely non-relativistic limit ($v\simeq0$), $p_{1}\simeq
p_{1}^{\prime}\simeq(p_{2}+p_{2}^{\prime})/2\simeq(M,\mathbf{0})$. \ The
invariant Mandelstam parameters are%

\begin{subequations}
\begin{align}
s  &  =(p_{1}+p_{1}^{\prime})^{2}=(p_{2}+p_{2}^{\prime})^{2}\simeq4M^{2}\\
t  &  =(-p_{1}+p_{2})^{2}=(p_{1}^{\prime}-p_{2}^{\prime})^{2}\simeq-M^{2}\\
u  &  =(-p_{1}^{\prime}+p_{2})^{2}=(p_{1}-p_{2}^{\prime})^{2}\simeq-M^{2}%
\end{align}

The formula reduces to%

\end{subequations}
\begin{equation}
\mathcal{M}_{B}=g_{Y}^{2}\widetilde{Y}_{L}^{2}\overline{u}(p_{2})\frac
{({p}_{2}-{p}_{2}^{\prime})^{\mu}\gamma^{\nu}+({p}_{2}-{p}_{2}^{\prime})^{\nu
}\gamma^{\mu}}{M^{2}+\widetilde{M}^{2}}P_{L}v(p_{2}^{\prime})\epsilon_{\mu
}(p_{1})\epsilon_{\nu}(p_{1}^{\prime})\ +\text{RH}%
\end{equation}

The consistent annihilation rate in $d=4-2\epsilon$ dimensions is%

\begin{equation}
\sigma_{B}v=\frac{(1-\epsilon)\Gamma(1-\epsilon)}{\Gamma(2-2\epsilon)}\left(
\frac{4\pi\mu^{2}}{4M^{2}}\right)  ^{\epsilon}\frac{2g_{Y}^{4}(\widetilde
{Y}_{L}^{4}+\widetilde{Y}_{R}^{4})N_{c}}{9\pi}\frac{M^{2}}{\left(
M^{2}+\widetilde{M}^{2}\right)  ^{2}}%
\end{equation}

where $N_{c}$ is number of colors, $v$ denotes the relative velocity between
spin-1 dark matter candidate pair, $\mu$ is an arbitrary mass scale and
$\sigma_{B}$ is the Born cross section$.$

\section{Next-to-leading order calculation}

The one-loop amplitudes consist of the virtual corrections from the diagrams
(a),(b),(c),(d) of Fig.2. Adding up those contributions must be ultraviolet
finite. The amplitudes contain infrared divergences due to massless gluon
virtual exchange. The infrared should cancel exactly against the one present
in the gluon final state radiation, Fig.3(a),(b),(c). The method is to use
dimensional regularization in $d=4-2\epsilon$ dimensions with massless
on-shell quarks to regularize both types of divergences, UV and IR
divergences. The individual diagrams are calculated in the renormalizable
Feynman gauge (the gauge parameter, $\xi=1$), which provides the gluon
propagator. \ All new particles involved in the calculation are set up to have
the same mass as $B^{\prime}$.

\subsection{Virtual corrections}%

\begin{figure}
[ptb]
\begin{center}
\includegraphics[
trim=0.000000in 0.000000in 0.000000in -0.108763in,
height=4.8677cm,
width=7.5278cm
]%
{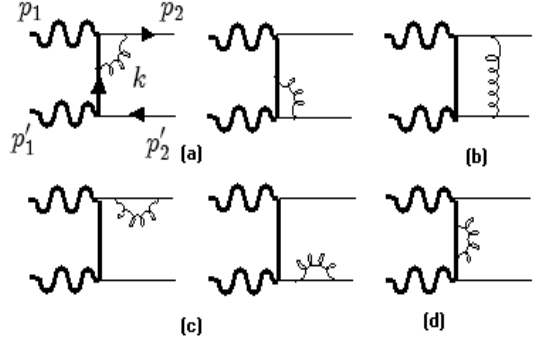}%
\caption{The Feynman diagrams which contribute to the next-to-leading order
(NLO) QCD virtual radiative correction($B^{\prime}B^{\prime}\rightarrow
q\overline{q}$). The crossed ($u-$channel) diagrams are not displayed. The
bold wavy lines represent the DM candidate gauge bosons ($B^{\prime}$), the
bold solid lines represent the heavy quarks (the $Z_{2}$-odd partners of the
outgoing quarks), curly lines represent the gluons and the light solid lines
correspond to the quarks. }%
\label{fig2}%
\end{center}
\end{figure}

All the virtual corrections could be expressed by two form
factors\footnote{Two more form factors are possible, which are related to the
tensors $({\not p  }_{2}\pm{\not p  }_{2}^{\prime})g^{\mu\nu}$. \ However both
do not give any contribution for the massless outgoing particles.} related to
tensors, $({p}_{2}-{p}_{2}^{\prime})^{\mu}\gamma^{\nu}+({p}_{2}-{p}%
_{2}^{\prime})^{\nu}\gamma^{\mu}$ and $({p}_{2}+{p}_{2}^{\prime})^{\mu}%
\gamma^{\nu}+({p}_{2}+{p}_{2}^{\prime})^{\nu}\gamma^{\mu}$, for the massless
outgoing particles in static limit. However $({p}_{2}-{p}_{2}^{\prime})^{\mu
}\gamma^{\nu}+({p}_{2}-{p}_{2}^{\prime})^{\nu}\gamma^{\mu}$ only survives.
${p}_{2}+{p}_{2}^{\prime}$ is a timelike vector and $B^{\prime}$ polarization
vectors, $\epsilon_{\mu}(p_{1})$ and $\epsilon_{\nu}(p_{1}^{\prime})$, are
spacelike. The tensor $({p}_{2}+{p}_{2}^{\prime})^{\mu}\gamma^{\nu}+({p}%
_{2}+{p}_{2}^{\prime})^{\nu}\gamma^{\mu}$ disappears as contracting the
$B^{\prime}$ polarization vectors. So the virtual corrections are expressed
with only one form factor, which is the coefficient of the Born amplitude.%

\begin{equation}
\mathcal{M}_{1}=F_{B^{\prime}}\mathcal{M}_{B}%
\end{equation}

We use the conventional approach of Feynman parameters to calculate the
corrections. The virtual corrections are simplified in the common integral
with the shifted momentum and Feynman parameters. The infrared divergences
appear in the Feynman parameter integrations.%

\begin{equation}
\int\frac{d^{d}\ell dx_{i}}{(2\pi)^{d}}\delta(%
{\textstyle\sum}
x_{i}-1)\frac{(\ell^{2})^{r}}{(\ell^{2}-C)^{m}}=\frac{i(-1)^{r-m}}%
{(4\pi)^{d/2}}\frac{\Gamma(r+d/2)\Gamma(m-r-d/2)}{\Gamma(d/2)\Gamma(m)}\int
dx_{i}\delta\left(
{\textstyle\sum}
x_{i}-1\right)  C^{r-m+d/2}%
\end{equation}

The amplitude of the diagrams in Fig.2(a) is%

\begin{equation}
\mathcal{M}_{1,L}^{(a)}=-g_{Y}^{2}\widetilde{Y}_{L}^{2}\overline{u}%
(p_{2})\left[  \Gamma_{(a)}^{\mu}\left(  \frac{{\not p  }_{2}-{\not p  }%
_{1}+M}{t-M^{2}}\right)  \gamma^{\nu}+\gamma^{\mu}\left(  \frac{{\not p  }%
_{2}-{\not p  }_{1}+M}{t-M^{2}}\right)  \widetilde{\Gamma}_{(a)}^{\nu}%
+\mu\leftrightarrow\nu\right]  P_{L}v(p_{2}^{\prime})\epsilon_{\mu}%
(p_{1})\epsilon_{\nu}(p_{1}^{\prime})
\end{equation}

$\Gamma_{(a)}^{\mu}$ and $\widetilde{\Gamma}_{(a)}^{\nu}$ are the vertex
corrections in $d-$dimensions, which are given by%

\begin{subequations}
\begin{align}
\Gamma_{(a)}^{\mu}  &  =-ig_{s}^{2}C_{F}%
{\displaystyle\int}
\frac{d^{d}k}{(2\pi)^{d}}\frac{\gamma^{\rho}({\not p  _{2}}+{\not k  }%
)\gamma^{\mu}({\not p  _{2}-\not p  _{1}}+{\not k  }+M)\gamma_{\rho}}%
{k^{2}(p_{2}+k)^{2}\left(  (p_{2}-p_{1}+k)^{2}-M^{2}\right)  }\\
\widetilde{\Gamma}_{(a)}^{\nu}  &  =-ig_{s}^{2}C_{F}%
{\displaystyle\int}
\frac{d^{d}k}{(2\pi)^{d}}\frac{\gamma^{\rho}({\not p  }^{\prime}{_{1}-\not p
_{2}}+{\not k  }+M)\gamma^{\nu}(-{\not p  }^{\prime}{_{2}}+{\not k  }%
)\gamma_{\rho}}{k^{2}(-p_{2}^{\prime}+k)^{2}\left(  (p_{1}^{\prime}%
-p_{2}^{\prime}+k)^{2}-M^{2}\right)  }%
\end{align}
where $g_{s}$ is the strong coupling and $C_{F}$ is the Casmir operator of the
fundamental representation in the color group. The Lorentz indices $\mu$ and
$\nu$ are just switched for the $u-$channel, since the invariant Mandelstam
parameters $t$ and $u$ are identical and the momenta, $p_{1}$ and
$p_{1}^{\prime},$ are timelike in the extremely non-relativistic
case($v\simeq0$).

The form factor of Fig.2(a) results in%

\end{subequations}
\begin{equation}
F_{B^{\prime}}^{(a)}=\frac{\alpha_{s}C_{F}}{2\pi}\left(  \frac{4\pi\mu^{2}%
}{4M^{2}}\right)  ^{\epsilon}\Gamma(1+\epsilon)\left(  \frac{1}{\epsilon_{UV}%
}+1+\log2\right)
\end{equation}
where the subscript UV implies the $1/\epsilon$ pole coming from UV divergence
and $\alpha_{s}(=g_{s}^{2}/4\pi)$ is QCD fine structure constant.

Calculation\footnote{The analogous calculation for the box diagram is in
Ref.\cite{khagi} for the different phenomenology.} of Fig.2(b) requires
careful and tedious effort because of four propagators in the loop and a three
folded integral over Feynman parameters. The scattering amplitude is%

\begin{equation}
\mathcal{M}_{1,L}^{(b)}=-g_{Y}^{2}\widetilde{Y}_{L}^{2}\overline{u}%
(p_{2})\Gamma_{(b)}^{\mu\nu}P_{L}v(p_{2}^{\prime})\epsilon_{\mu}%
(p_{1})\epsilon_{\nu}(p_{1}^{\prime})
\end{equation}

$\Gamma_{(b)}^{\mu\nu}$ is the combined vertex correction of the $t-$ and
$u-$channel diagrams. It is given by%

\begin{equation}
\Gamma_{(b)}^{\mu\nu}=-ig_{s}^{2}C_{F}\int\frac{d^{d}k}{(2\pi)^{d}}%
\frac{\gamma^{\rho}({\not p  }_{2}+{\not k  )}\gamma^{\mu}({\not p
_{2}-\not p  _{1}}+{\not k  +M)}\gamma^{\nu}(-{\not p  }_{2}^{\prime}+{\not k
)}\gamma_{\rho}+\mu\longleftrightarrow\nu}{k^{2}(p_{2}+k)^{2}(-p_{2}^{\prime
}+k)^{2}((p_{2}-p_{1}+k)^{2}-M^{2})}%
\end{equation}

This integration could also be manipulated into the common form, Eq.(7),
however the integral has singularity in the euclidean region. The imaginary
parts are included to continue this integral in the euclidean region for the
positive $s$ and it results in a complex form factor.%

\begin{equation}
F_{B^{\prime}}^{(b)}=\frac{\alpha_{s}C_{F}}{2\pi}\left(  \frac{4\pi\mu^{2}%
}{4M^{2}}\right)  ^{\epsilon}\Gamma(1+\epsilon)\left(  -\frac{1}{\epsilon
_{IR}^{2}}-\frac{2}{\epsilon_{IR}}-\frac{14}{3}+\frac{2}{3}\log2+\frac
{2\pi^{2}}{3}+i\pi\left(  \frac{1}{\epsilon_{IR}}+3+\frac{\pi}{6}\right)
\right)
\end{equation}

The imaginary parts are not relevant for real radiative meseurements as a
consequence of the unitarity of the $S$-matrix:$S^{\dag}S=1$.

The contribution of Fig.2(c) comes from propagator corrections to on-shell
quark lines. For the massless quark lines, there is no contribution using
dimensional regularization since the same regulator is used. In the leading
order of $\epsilon$, it can be written as%

\begin{equation}
F_{B^{\prime}}^{(c)}=\frac{\alpha_{s}C_{F}}{2\pi}\left(  \frac{4\pi\mu^{2}%
}{4M^{2}}\right)  ^{\epsilon}\Gamma(1+\epsilon)\left(  -\frac{1}%
{2\epsilon_{UV}}+\frac{1}{2\epsilon_{IR}}\right)
\end{equation}

However the contribution of Fig.2(d) comes from off-shell heavy quark lines.
It produces a contribution without IR divergence.%

\begin{equation}
F_{B^{\prime}}^{(d)}=2\frac{d\Sigma_{2}}{d{\not p  }}|_{p^{2}=-M^{2}}%
=\frac{\alpha_{s}C_{F}}{2\pi}\left(  \frac{4\pi\mu^{2}}{4M^{2}}\right)
^{\epsilon}\Gamma(1+\epsilon)\left(  -\frac{1}{2\epsilon_{UV}}-1+\log2\right)
\end{equation}
where the factor of 2 comes from the two vertices and those give identical
contributions. The results show that UV divergences are exactly canceled.

Adding up all the virtual corrections, the QCD corrections are%

\begin{equation}
\delta_{QCD}^{(\operatorname*{virutal})}=2\operatorname{Re}(F_{B^{\prime}%
})=\frac{\alpha_{s}C_{F}}{\pi}\left(  \frac{4\pi\mu^{2}}{4M^{2}}\right)
^{\epsilon}\Gamma(1+\epsilon)\left(  -\frac{1}{\epsilon_{IR}^{2}}-\frac
{3}{2\epsilon_{IR}}+\frac{2\pi^{2}}{3}-\frac{14}{3}+\frac{8}{3}\log2\right)
\end{equation}

\subsection{\bigskip Real corrections}%

\begin{figure}
[ptb]
\begin{center}
\includegraphics[
trim=0.000000in 0.000000in -0.253466in 0.000000in,
height=2.5085cm,
width=7.027cm
]%
{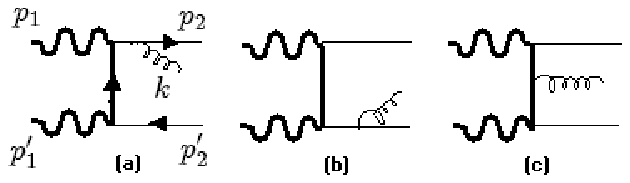}%
\caption{The Feynman diagrams which contribute to the next-to-leading order
(NLO) QCD real radiative correction($B^{\prime}B^{\prime}\rightarrow
gq\overline{q}$). The crossed ($u-$channel) diagrams are not displayed. The
bold wavy lines represent the DM candidate gauge bosons ($B^{\prime}$), the
bold solid lines represent the heavy quarks (the $Z_{2}$-odd partners of the
outgoing quarks), curly lines represent the gluons and the light solid lines
correspond to the quarks. }%
\label{fig3}%
\end{center}
\end{figure}

The real QCD correction appears in the ratio of annihilation rates for two and
three body final states. In an average over the polarization of the incoming
vector bosons and a sum over the spin and color of outgoing quarks and gluon,
the annihilation rate for three body final states is%

\begin{equation}
\sigma v=\frac{1}{4M^{2}}\cdot\frac{N_{C}}{9}%
{\textstyle\int}
d\Phi_{3}|\mathcal{M}_{a}+\mathcal{M}_{b}+\mathcal{M}_{c}|^{2}%
\end{equation}

$\mathcal{M}_{a},\mathcal{M}_{b}$ and $\mathcal{M}_{c}$ are the scattering
amplitudes corresponded to the diagrams (a),(b),(c) of Fig.3, and are given by%

\begin{subequations}
\begin{align}
\mathcal{M}_{a,L}  &  =-ig_{Y}^{2}g_{s}\widetilde{Y}_{L}^{2}T^{a}%
\frac{\overline{u}(p_{2})\gamma^{\rho}({\not p  }_{2}+{\not k  })\left(
\gamma^{\mu}{\not p  }_{a}\gamma^{\nu}+\gamma^{\nu}{\not p  }_{a}\gamma^{\mu
}\right)  P_{L}v(p_{2}^{\prime})\epsilon_{\mu}(p_{1})\epsilon_{\nu}%
(p_{1}^{\prime})\epsilon_{\rho}^{\ast a}(k)}{(p_{2}+k)^{2}(p_{a}^{2}-M^{2})}\\
\text{\ }\mathcal{M}_{b,L}  &  =-ig_{Y}^{2}g_{s}\widetilde{Y}_{L}^{2}%
T^{a}\frac{\overline{u}(p_{2})\left(  \gamma^{\mu}{\not p  }_{b}\gamma^{\nu
}+\gamma^{\nu}{\not p  }_{b}\gamma^{\mu}\right)  (-{\not p  }_{2}^{\prime
}-{\not k  })\gamma^{\rho}P_{L}v(p_{2}^{\prime})\epsilon_{\mu}(p_{1}%
)\epsilon_{\nu}(p_{1}^{\prime})\epsilon_{\rho}^{\ast a}(k)}{(-p_{2}^{\prime
}-k)^{2}(p_{b}^{2}-M^{2})}\\
\mathcal{M}_{c,L}  &  =-ig_{Y}^{2}g_{s}\widetilde{Y}_{L}^{2}T^{a}%
\frac{\overline{u}(p_{2})\left[  \gamma^{\mu}{\not p  }_{b}\gamma^{\rho
}{\not p  }_{a}\gamma^{\nu}+M^{2}\gamma^{\mu}\gamma^{\rho}\gamma^{\nu}%
+\mu\leftrightarrow\nu\right]  P_{L}v(p_{2}^{\prime})\epsilon_{\mu}%
(p_{1})\epsilon_{\nu}(p_{1}^{\prime})\epsilon_{\rho}^{\ast a}(k)}{(p_{a}%
^{2}-M^{2})(p_{b}^{2}-M^{2})}%
\end{align}
where $T^{a}$ is the QCD generator and $p_{a}=p_{1}^{\prime}-p_{2}^{\prime
}=-p_{1}+p_{2}+k,p_{b}=-p_{1}+p_{2}=p_{1}^{\prime}-p_{2}^{\prime}-k$ in
shorthand where $p_{1}$and $p_{1}^{\prime}$ are momenta of the incoming vector
bosons, $p_{2}$ and $p_{2}^{\prime}$ for the outgoing quarks and $k$ for the
radiated gluon.

$p_{1}\simeq p_{1}^{\prime}\simeq(p_{2}+p_{2}^{\prime}+k)/2\simeq
(M,\mathbf{0})$ in the non-relativistic limit and the numerator of
$\mathcal{M}_{c}$ is allowed to express by the momenta, $p=p_{2}-p_{2}%
^{\prime}$ and $k$ on the diagrammetic symmetry.%

\end{subequations}
\begin{equation}
\mathcal{N}_{c}\sim p^{\rho}(p^{\mu}\gamma^{\nu}+p^{\nu}\gamma^{\mu
})+ip_{\sigma}k_{\delta}\gamma^{5}\epsilon^{\sigma\rho\delta\lambda
}(g_{\lambda}^{\mu}\gamma^{\nu}+g_{\lambda}^{\nu}\gamma^{\mu}-g^{\mu\nu}%
\gamma_{\lambda})+(p_{2}\cdot p_{2}^{\prime}+2M^{2})(g^{\mu\rho}\gamma^{\nu
}+g^{\nu\rho}\gamma^{\mu}-g^{\mu\nu}\gamma^{\rho})
\end{equation}

The relation, $\epsilon_{\rho}(p)\epsilon_{\sigma}^{\ast}(p^{\prime}%
)=-g_{\rho\sigma}$, is adopted to calculate the squared amplitude for both
incoming massive spin-1 dark matter candidates in Feynman gauge, since the
polarization vectors $\epsilon_{\mu}(p_{1}),\epsilon_{\nu}(p_{1}^{\prime}%
)$\ are spacelike. The soft and collinear singularities appear in the squared
amplitudes,
$\vert$%
$\mathcal{M}_{a}|^{2},$%
$\vert$%
$\mathcal{M}_{b}|^{2}$ and $2\operatorname{Re}(\mathcal{M}_{a}^{\ast
}\mathcal{M}_{b})$. $1/\epsilon$ poles are canceled out in 2$\operatorname{Re}%
(\mathcal{M}_{a}^{\ast}\mathcal{M}_{c})$ and 2$\operatorname{Re}%
(\mathcal{M}_{b}^{\ast}\mathcal{M}_{c})$ as combining the numerator and
denominator, and $\left\vert \mathcal{M}_{c}\right\vert ^{2}$ is totally free
of IR divergences.

The new dimensionless parameters are introduced for the phase space
integration by energy-momentum conservation.%

\begin{equation}
x_{1}=\frac{p_{2}\cdot k}{2M^{2}},\qquad x_{2}=\frac{p_{2}^{\prime}\cdot
k}{2M^{2}},\qquad x_{3}=\frac{p_{2}\cdot p_{2}^{\prime}}{2M^{2}}%
\end{equation}

$1/\epsilon$ poles appear at $x_{1}=0$ and $x_{2}=0.$ The Lorentz-invariant
three body phase space takes the form with the new parameters.%

\begin{equation}%
{\textstyle\int}
d\Phi_{3}=\frac{4M^{2}}{(4\pi)^{3}\Gamma(2-2\epsilon)}\left(  \frac{4\pi
\mu^{2}}{4M^{2}}\right)  ^{2\epsilon}%
{\textstyle\int}
dx_{1}dx_{2}dx_{3}(x_{1}x_{2}x_{3})^{-\epsilon}\delta(%
{\textstyle\sum}
x_{i}-1)
\end{equation}

The overall real correction gives%

\begin{equation}
\delta_{QCD}^{(\operatorname{real})}=\frac{\alpha_{s}C_{F}}{\pi}\left(
\frac{4\pi\mu^{2}}{4M^{2}}\right)  ^{\epsilon}\Gamma(1+\epsilon)\left(
\frac{1}{\epsilon_{IR}^{2}}+\frac{3}{2\epsilon_{IR}}+\frac{39}{4}-\frac{7}%
{6}\pi^{2}\right)
\end{equation}

When all the contributions\footnote{The corrections can also be approached by
the optical theorem to acquire the cross section. But eventually both are
identical and we would have the same QCD correction.} are added, the infrared
are exactly canceled and the finite result is%

\begin{equation}
\delta_{QCD}=\frac{\alpha_{s}C_{F}}{\pi}\left(  \frac{61}{12}+\frac{8}{3}%
\log2-\frac{\pi^{2}}{2}\right)  \text{ }%
\end{equation}

The correction is about 8$\%$ enhancement for $C_{F}=4/3$ and $\alpha_{s}%
($1TeV$)=0.09$, and the finiteness of this correction implies that there is no
divergence by degeneracy.

\subsection{The corrections by the heavy gluon}

If the heavy $Z_{2}$-odd of the usual gluon is not far above $M$, its
contribution to the $B^{\prime}$ pair annihilation could be important. The
relevant Lagrangian of this heavy gluon field $G_{\mu}^{\prime a}$ is%

\begin{equation}
\mathcal{L}\supset-g_{s}\overline{\widetilde{q}}\gamma^{\mu}T^{a}G_{\mu
}^{^{\prime}a}q+h.c.
\end{equation}

For being able to carry through analytical calculation, we set the mass of
this heavy gluon field $G_{\mu}^{\prime a}$ equal to the $B^{\prime}$ mass $M$
in our calculation. Such a simplification is valid at least in UED. The
corresponding Feynman diagrams are similarly given in Fig. 2, except that we
need to interchange the bold quark line and the light quark line in the loop
as well as the usual gluon is replaced by the heavy gluon. Note that there is
no IR divergence for a massive gluon, and the UV divergence cancels among
diagrams. As a single heavy gluon cannot be produced by $Z_{2}$-symmetry, the
diagrams in Fig. 3 are all absent and it results in the improved QCD
correction. The QCD corrections by the heavy gluon can also be calculated
fully analytically and the overall correction is%

\begin{equation}
\delta_{QCD}^{(\operatorname*{gluon})}=\frac{\alpha_{s}C_{F}}{\pi}\left(
-5+\frac{9\pi^{2}}{8}-10\log2+\frac{7\pi}{6\sqrt{3}}\right)  \text{ }%
\end{equation}

The correction is about $5\%$ enhancement and is comparable to massless gluon correction.

\section{Application to relic abundance}%

\begin{figure}
[t]
\begin{center}
\includegraphics[
trim=0.000000in 0.000000in 0.000000in -0.094793in,
height=2.4794in,
width=3.4835in
]%
{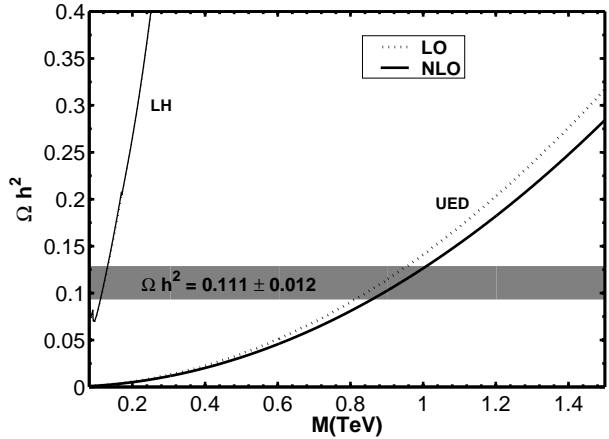}%
\caption{Prediction for relic abundance as a function of the WIMP mass. NLO
QCD radiative corrections are considered on the solid lines. The horizonal
band denotes $\Omega$h$^{2}$=0.111$\pm$0.012 and defines the WIMP mass window.
The WIMP mass in the window is shifted about 50 GeV for UED with NLO QCD
correction.}%
\label{fig4}%
\end{center}
\end{figure}

The spin-1 dark matter candidate $B^{\prime}$, which is the $Z_{2}$-odd
partner of the hypercharge gauge boson $B$, has been an attractive dark matter
candidate in UED and LH, but both models were extended in different ways. The
spatial dimensions are enlarged for UED, otherwise the symmetry groups are
enlarged for LH in 4-dimensions. So the masses of the new particles are scaled
by the extra dimension of size $R$ compactified on an $S^{1}/Z_{2}$ orbifold
for UED and the enlarged global symmetry breaking scale $f$ \ for LH. Since
the new particles are in the different symmetry group structure, the different
gauge charges are assigned to the fermions ($q\widetilde{q}B^{\prime}$
coupling) and it causes them to induce the different phenomenological
analyses. We consider relic abundance with our QCD correction, but the
detailed analyses are not mentioned since those were well studied in the
leading order in the other articles\cite{geraldine, andreas}.

Assuming that the spin-1 $B^{\prime}$ accounts for all DM relic abundance, one
can constrain its mass $M$ based on the pair annihilation rate in the early
universe. In UED models, the annihilation channel into a fermion pair
$B^{\prime}B^{\prime}\rightarrow f\overline{f}$ dominates because of sizable
couplings of the gauge charges $\widetilde{Y}_{L/R}$. The measured relic
abundance prefers $M$ to be about TeV. On the contrary, the L$^{2}$H model has
rather small couplings of the gauge charges $\widetilde{Y}_{L}=\frac{1}{10}%
$,$\widetilde{Y}_{R}=0$, and the relic abundance only requires $M$ of order of
100 GeV. The detailed quantitative analyses for relic abundance by this type
of WIMP annihilations can be found in Ref.\cite{geraldine} for UED and
Ref.\cite{andreas} for LH. Fig.4 shows the prediction for relic abundance as a
function of WIMP mass with the present WMAP precision \cite{dnspergel}. The
WIMP mass in the window is shifted about 50 GeV compared to the leading order
for UED\footnote{It was truncated to the first level Kaluza Klein (KK) mode
for our calculation and UED with such truncation is renormalizable though the
theory is not renormalizable.}, but it shows no difference from the leading
order by the negligible annihilation fractions into quarks for LH . \ 

\section{Conclusion}

The next-to-leading order (NLO) QCD corrections are calculated for
pair-annihilation of spin-1 bosonic dark matter by dimensionally regularizing
both ultraviolet and infrared singularities in the non-relativistic limit. The
order $\alpha_{s}$ correction amounts to about $8\%$ and can enhance to
$13\%,$ when including heavy gluon. The NLO QCD correction could give the
sizable shift to the DM mass constrained by relic density measurements.

\begin{acknowledgments}
The author would like to thank Professor Wai-Yee Keung for reading the
manuscript carefully.
\end{acknowledgments}

\appendix*

\section{Useful formulas and identities}

The Dirac algebra in $d=4-2\epsilon$ dimensions are listed in Ref.\cite{wjmar}
and they could be induced by the anticommutator $\left\{  \gamma^{\mu}%
,\gamma^{\nu}\right\}  =2g^{\mu\nu}$ and the identity $\gamma^{\mu}\gamma
_{\mu}=d$. The regulator $\epsilon$ is omitted on trace algebra Tr$\left[
\text{I}\right]  =2^{d/2}$ for two and three body cross section calculations.

$1/\epsilon_{IR}$ poles are extracted by the partial integrations on the loop
calculations in case of a need to extract the poles and the remnants are
expanded in ordinary Taylor series with respect to $\epsilon$. The three
folded parametric integrals for the box diagram are calculated by Feynman
parameter properties, since one of them is not symmetric with the others.%

\begin{equation}
\int_{0}^{1}dx\int_{0}^{1-x}dyf(x,y)=\int_{0}^{1}dy\int_{0}^{1-y}dxf(x,y)
\end{equation}

The complicated integrations are avoided and number of integrations is reduced
with this property.

The useful identity to split the singular and non-singular parts for the real
corrections is%
\begin{equation}
\frac{1}{x(1-x)}=\frac{1}{x}+\frac{1}{1-x}%
\end{equation}

This identity can be extended to the higher powers of the variables and will
simplify the calculations since the regulator $\epsilon$ can be dropped for
non-singular parts.

Some of the parametric integrals contain the dilogarithm (or spence) function.%

\begin{equation}
\text{Li}_{2}(x)\equiv-\int_{0}^{1}dt\frac{\log(1-xt)}{t}%
\end{equation}

The values which appear in our calculations are%

\begin{subequations}
\begin{align}
\text{Li}_{2}(1)  &  =\frac{\pi^{2}}{6}\text{ }\\
\text{\ Li}_{2}(-1)  &  =-\frac{\pi^{2}}{12}\text{ }\\
\text{\ Li}_{2}\left(  \frac{1}{2}\right)   &  =\frac{\pi^{2}}{12}-\frac
{\log^{2}2}{2}%
\end{align}
\newline

The parametric integrals which appeared in the box diagrams reduce to the
incomplete beta function.%

\end{subequations}
\begin{equation}
B_{1/2}(m,n)\equiv\int_{0}^{1/2}dxx^{m-1}\left(  1-x\right)  ^{n-1}%
\end{equation}

\begin{subequations}
\begin{align}
B_{1/2}(m,n)+B_{1/2}(n,m)  &  =B(m,n)\\
B_{1/2}(m,m)  &  =\frac{1}{2}B(m,m)\\
B_{1/2}(m,m+1)  &  =\frac{1}{2}\left[  B(m,m+1)+\frac{2^{-2m}}{m}\right] \\
B_{1/2}(m,m+2)  &  =\frac{1}{2}\left[  B(m,m+2)+\frac{2^{-2m}}{m}\right]
\end{align}

where%

\end{subequations}
\begin{equation}
B(m,n)=\frac{\Gamma(m)\Gamma(n)}{\Gamma(m+n)}%
\end{equation}

For integer $n$, the Taylor expansion is adopted with respect to $\epsilon$.%

\begin{subequations}
\begin{align}
B_{z}(-\epsilon,0)  &  =1+\epsilon\log(1-z)-\epsilon^{2}\text{Li}_{2}(z)\\
B_{z}(-\epsilon,1)  &  =1+\epsilon\left(  -\frac{z}{1-z}+\log(1-z)\right)
+\epsilon^{2}\left(  \log(1-z)-\text{Li}_{2}(z)\right)
\end{align}

with $z=1/2.$

The parametric integrations involved the heavy gluon loops\ are%

\end{subequations}
\begin{align*}
\int_{0}^{1}dx\log(x^{2}-x+1)  &  =2\left(  \frac{\sqrt{3}}{6}\pi-1\right) \\
\int_{0}^{1}dx(x^{2}-x)\log(x^{2}-x+1)  &  =\frac{17}{18}-\frac{\sqrt{3}}%
{6}\pi\\
\int_{0}^{1}dx\frac{x^{2}-x}{x^{2}-x+1}  &  =-1+\frac{2}{3\sqrt{3}}\pi
\end{align*}

\end{document}